**Expansion and one-range addition theorems for complete orthonormal sets of spinor wave functions and Slater spinor orbitals of arbitrary half-integral spin in position, momentum and four-dimensional spaces**

I.I. Guseinov

*Department of Physics, Faculty of Arts and Sciences,*

*Onsekiz Mart University, Çanakkale, Turkey*

**Abstract**

The analytical relations in position, momentum and four-dimensional spaces are established for the expansion and one-range addition theorems of relativistic complete orthonormal sets of exponential type spinor wave functions and Slater spinor orbitals of arbitrary half-integral spin. These theorems are expressed through the corresponding nonrelativistic expansion and one-range addition theorems of the spin-0 particles introduced by the author. The expansion and one-range addition theorems derived are especially useful for the computation of multicenter integrals over exponential type spinor orbitals arising in the generalized relativistic Dirac-Hartree-Fock-Roothaan theory when the position, momentum and four-dimensional spaces are employed.

## 1. Introduction

The solutions of the Dirac equation for hydrogen-like systems play a significant role in theory and application to relativistic quantum mechanics of atoms, molecules and nuclei. However, the relativistic hydrogen-like position orbitals and their extensions to momentum and four-dimensional spaces cannot be used as basis sets because they are not complete unless the continuum is included [1-4]. In Ref. [5] we have constructed in position, momentum and four-dimensional spaces the complete orthonormal sets of two- and four-component relativistic spinor wave functions based on the use of complete othonormal sets of nonrelativistic orbitals. By the use of this method, in a previous work [6], we introduced the new complete orthonormal sets of relativistic $\Psi^{\alpha s}$-exponential type spinor orbitals ($\Psi^{\alpha s}$-ETSO) and $X^{s}$-Slater type spinor orbitals ($X^{s}$-STSO) for particles with arbitrary half-integral spin in position, momentum and four-dimensional spaces through the corresponding nonrelativistic $\psi^{\alpha}$-exponential type orbitals ($\psi^{\alpha}$-ETO) [7] and $\chi$-Slater type orbitals ($\chi$-STO). The elaboration of algorithm for the solution of generalized Dirac equations [8] in linear combination of atomic spinor orbitals (LCASO) approach necessitates progress in the development of theory for one-range addition theorems of spinor orbitals of multiple order.

Addition theorems play a more and more important role in nonrelativistic and relativistic atomic and molecular electronic structure calculations [9]. Two fundamentally different types of addition theorems occur in the literature. The first type of the addition theorems has the two-range form of Laplace expansion for the Coulomb potential. There is second class of addition theorems which can be constructed by expanding a function located at a center $a$ in terms of a complete orthonormal set located at a center $b$. The use of one-range addition theorems in electronic structure calculations would be highly desirable since they are capable of producing much better approximations than the two-range addition theorems. In Refs.[10-13] we have developed the method for constructing in position, momentum and four-dimensional spaces the one-range addition theorems of complete orthonormal sets of nonrelativistic $\psi^{\alpha}$-ETO and $\chi$-STO. The aim of this work is to derive the relevant expansion and one-range addition theorems of complete orthonormal sets of relativistic $\Psi^{\alpha s}$-ETSO and $X^{s}$-STSO in position, momentum and four-dimensional spaces through the corresponding theorems for nonrelativistic orbitals $\psi^{\alpha}$-ETO and $\chi$-STO. These theorems might be useful for the calculation of multicenter integrals which appear in relativistic MO LCASO theory of arbitrary half-integral spin particles when the spinor orbitals basis sets in position, momentum and four-dimensional spaces are employed.

## 2. Definitions and basic formulas

In order to derive the expansion and one-range addition theorems for 2(2s+1)-component spinor orbitals in position, momentum and four-dimensional spaces we use the following definitions:

Complete orthonormal sets of nonrelativistic orbitals

$$k^{\alpha}_{nlm}(\zeta,\vec{x}) \equiv \psi^{\alpha}_{nlm}(\zeta,\vec{r}), \phi^{\alpha}_{nlm}(\zeta,\vec{k}), z^{\alpha}_{nlm}(\zeta,\vec{\omega}_k) \tag{1}$$

$$\bar{k}^{\alpha}_{nlm}(\zeta,\vec{x}) \equiv \bar{\psi}^{\alpha}_{nlm}(\zeta,\vec{r}), \bar{\phi}^{\alpha}_{nlm}(\zeta,\vec{k}), \bar{z}^{\alpha}_{nlm}(\zeta,\vec{\omega}_k), \tag{2}$$

Slater type nonrelativistic spinor orbitals

$$k_{nlm}(\zeta,\vec{x}) \equiv \chi_{nlm}(\zeta,\vec{r}), u_{nlm}(\zeta,\vec{k}), v_{nlm}(\zeta,\vec{\omega}_k), \tag{3}$$

Complete orthonormal sets of 2(2s+1)-component relativistic spinor orbitals

$${}^{t}K^{\alpha s}_{nljm_j}(\zeta,\vec{x}) \equiv {}^{t}\Psi^{\alpha s}_{nljm_j}(\zeta,\vec{r}), {}^{t}\Phi^{\alpha s}_{nljm_j}(\zeta,\vec{k}), {}^{t}Z^{\alpha s}_{nljm_j}(\zeta,\vec{\omega}_k) \tag{4a}$$

$${}^{t}\boldsymbol{K}^{\alpha s}_{nl_t jm_j}(\zeta,\vec{x}) \equiv {}^{t}\boldsymbol{\Psi}^{\alpha s}_{nl_t jm_j}(\zeta,\vec{r}), {}^{t}\boldsymbol{\Phi}^{\alpha s}_{nl_t jm_j}(\zeta,\vec{k}), {}^{t}\boldsymbol{Z}^{\alpha s}_{nl_t jm_j}(\zeta,\vec{\omega}_k) \tag{4b}$$

$${}^{t}\bar{K}^{\alpha s}_{nljm_j}(\zeta,\vec{x}) \equiv {}^{t}\bar{\Psi}^{\alpha s}_{nljm_j}(\zeta,\vec{r}), {}^{t}\bar{\Phi}^{\alpha s}_{nljm_j}(\zeta,\vec{k}), {}^{t}\bar{Z}^{\alpha s}_{nljm_j}(\zeta,\vec{\omega}_k) \tag{5a}$$

$${}^{t}\bar{\boldsymbol{K}}^{\alpha s}_{nl_t jm_j}(\zeta,\vec{x}) \equiv {}^{t}\bar{\boldsymbol{\Psi}}^{\alpha s}_{nl_t jm_j}(\zeta,\vec{r}), {}^{t}\bar{\boldsymbol{\Phi}}^{\alpha s}_{nl_t jm_j}(\zeta,\vec{k}), {}^{t}\bar{\boldsymbol{Z}}^{\alpha s}_{nl_t jm_j}(\zeta,\vec{\omega}_k), \tag{5b}$$

Slater type 2(2s+1)-component relativistic spinor orbitals

$${}^{t}K^{s}_{nljm_j}(\zeta,\vec{x}) \equiv {}^{t}X^{s}_{nljm_j}(\zeta,\vec{r}), {}^{t}U^{s}_{nljm_j}(\zeta,\vec{k}), {}^{t}V^{s}_{nljm_j}(\zeta,\vec{\omega}_k) \tag{6a}$$

$${}^{t}\boldsymbol{K}^{s}_{nl_t jm_j}(\zeta,\vec{x}) \equiv {}^{t}\boldsymbol{X}^{s}_{nl_t jm_j}(\zeta,\vec{r}), {}^{t}\boldsymbol{U}^{s}_{nl_t jm_j}(\zeta,\vec{k}), {}^{t}\boldsymbol{V}^{s}_{nl_t jm_j}(\zeta,\vec{\omega}_k), \tag{6b}$$

where $\vec{x} \equiv \vec{r}, \vec{k}, \vec{\omega}_k$ and $\omega_k \equiv \beta_k \theta_k \varphi_k$.

See Refs.[6] and [14-15] for the exact definition of quantities occurring in Eqs (1)-(6).

We shall also use the following formulas for 2(2s+1)-component spinor orbitals through the independent sets of two-component spinors defined as a product of complete orthonormal sets of radial parts of nonrelativistic scalar $\psi^{\alpha}$ -ETO and modified Clebsch-Gordan coefficients appearing in two-component tensor spherical harmonics (see Refs.[6] and [14-15]):

for $K^{\alpha s} - ETSO$

$${}^{t}K^{\alpha s}_{nljm_j}(\zeta,\vec{x}) = N_{nl_t} \begin{bmatrix} {}^{t}K^{\alpha s0}_{nljm_j}(\zeta,\vec{x}) \\ {}^{t}K^{\alpha s2}_{nljm_j}(\zeta,\vec{x}) \\ \vdots \\ {}^{t}K^{\alpha s,2s-1}_{nljm_j}(\zeta,\vec{x}) \\ {}^{t}\boldsymbol{K}^{\alpha s,2s-1}_{nl_t jm_j}(\zeta,\vec{x}) \\ \vdots \\ {}^{t}\boldsymbol{K}^{\alpha s2}_{nl_t jm_j}(\zeta,\vec{x}) \\ {}^{t}\boldsymbol{K}^{\alpha s0}_{nl_t jm_j}(\zeta,\vec{x}) \end{bmatrix} \tag{7a}$$

$$ {}^{t}K^{\alpha s\lambda}_{nljm_j}\left(\zeta,\vec{x}\right)=\begin{bmatrix}\eta_t\,{}^{t}a^{ls}_{jm_j}(\lambda)k^{\alpha}_{nlm(\lambda)}\left(\zeta,\vec{x}\right)\\ -\eta_t\,{}^{t}a^{ls}_{jm_j}(\lambda+1)k^{\alpha}_{nlm(\lambda+1)}\left(\zeta,\vec{x}\right)\end{bmatrix} \tag{7b} $$

$$ {}^{t}\boldsymbol{K}^{\alpha s\lambda}_{nl_tjm_j}\left(\zeta,\vec{x}\right)=\begin{bmatrix}-i\,{}^{t}a^{l_ts}_{jm_j}(2s-\lambda)k^{\alpha}_{nl_tm(\lambda)}\left(\zeta,\vec{x}\right)\\ -i\,{}^{t}a^{l_ts}_{jm_j}(2s-(\lambda+1))k^{\alpha}_{nl_tm(\lambda+1)}\left(\zeta,\vec{x}\right)\end{bmatrix} \tag{7c} $$

for $\bar{K}^{\alpha s}-ETSO$

$$ {}^{t}\bar{K}^{\alpha s}_{nljm_j}\left(\zeta,\vec{x}\right)=N_{nl_t}\begin{bmatrix}{}^{t}\bar{K}^{\alpha s0}_{nljm_j}\left(\zeta,\vec{x}\right)\\ {}^{t}\bar{K}^{\alpha s2}_{nljm_j}\left(\zeta,\vec{x}\right)\\ \vdots\\ {}^{t}\bar{K}^{\alpha s,2s-1}_{nljm_j}\left(\zeta,\vec{x}\right)\\ {}^{t}\bar{\boldsymbol{K}}^{\alpha s,2s-1}_{nl_tjm_j}\left(\zeta,\vec{x}\right)\\ \vdots\\ {}^{t}\bar{\boldsymbol{K}}^{\alpha s2}_{nl_tjm_j}\left(\zeta,\vec{x}\right)\\ {}^{t}\bar{\boldsymbol{K}}^{\alpha s0}_{nl_tjm_j}\left(\zeta,\vec{x}\right)\end{bmatrix} \tag{8a} $$

$$ {}^{t}\bar{K}^{\alpha s\lambda}_{nljm_j}\left(\zeta,\vec{x}\right)=\begin{bmatrix}\eta_t\,{}^{t}a^{ls}_{jm_j}(\lambda)\bar{k}^{\alpha}_{nlm(\lambda)}\left(\zeta,\vec{x}\right)\\ -\eta_t\,{}^{t}a^{ls}_{jm_j}(\lambda+1)\bar{k}^{\alpha}_{nlm(\lambda+1)}\left(\zeta,\vec{x}\right)\end{bmatrix} \tag{8b} $$

$$ {}^{t}\bar{\boldsymbol{K}}^{\alpha s\lambda}_{nl_tjm_j}\left(\zeta,\vec{x}\right)=\begin{bmatrix}-i\,{}^{t}a^{l_ts}_{jm_j}(2s-\lambda)\bar{k}^{\alpha}_{nl_tm(\lambda)}\left(\zeta,\vec{x}\right)\\ -i\,{}^{t}a^{l_ts}_{jm_j}(2s-(\lambda+1))\bar{k}^{\alpha}_{nl_tm(\lambda+1)}\left(\zeta,\vec{x}\right)\end{bmatrix} \tag{8c} $$

for $K^{s}-STSO$

$$ {}^{t}K^{s}_{nljm_j}\left(\zeta,\vec{x}\right)=N_{nl_t}\begin{bmatrix}{}^{t}K^{s0}_{nljm_j}\left(\zeta,\vec{x}\right)\\ {}^{t}K^{s2}_{nljm_j}\left(\zeta,\vec{x}\right)\\ \vdots\\ {}^{t}K^{s,2s-1}_{nljm_j}\left(\zeta,\vec{x}\right)\\ {}^{t}\boldsymbol{K}^{s,2s-1}_{nl_tjm_j}\left(\zeta,\vec{x}\right)\\ \vdots\\ {}^{t}\boldsymbol{K}^{s2}_{nl_tjm_j}\left(\zeta,\vec{x}\right)\\ {}^{t}\boldsymbol{K}^{s0}_{nl_tjm_j}\left(\zeta,\vec{x}\right)\end{bmatrix} \tag{9a} $$

$$ {}^{t}K^{s\lambda}_{nljm_j}\left(\zeta,\vec{x}\right)=\begin{bmatrix}\eta_t\,{}^{t}a^{ls}_{jm_j}(\lambda)k_{nlm(\lambda)}\left(\zeta,\vec{x}\right)\\ -\eta_t\,{}^{t}a^{ls}_{jm_j}(\lambda+1)k_{nlm(\lambda+1)}\left(\zeta,\vec{x}\right)\end{bmatrix} \tag{9b} $$

$$ {}^{t}\boldsymbol{K}^{s\lambda}_{nl_tjm_j}\left(\zeta,\vec{x}\right)=\begin{bmatrix}-i\,{}^{t}a^{l_ts}_{jm_j}(2s-\lambda)k_{nl_tm(\lambda)}\left(\zeta,\vec{x}\right)\\ -i\,{}^{t}a^{l_ts}_{jm_j}(2s-(\lambda+1))k_{nl_tm(\lambda+1)}\left(\zeta,\vec{x}\right)\end{bmatrix}, \tag{9c} $$

where $\lambda=0,2,...,2s-1$.

## 2. Expansion and one-range addition theorems for ETSO and STSO

With the derivation of expansion and one-range addition theorems for 2(2s+1)-component spinor orbitals in position, momentum and four-dimensional spaces, we use the method set out in previous papers [16-17] described for the nonrelativistic cases. Then, using Eqs. (7)-(9) and carrying through calculations analogous to those for the nonrelativistic basis sets we obtain the following relations in terms of nonrelativistic cases:

EXPANSION THEOREMS:

for ETSO

$$ {}^{t}K^{\alpha s^+}_{nljm_j}\left(\zeta,\vec{x}\right){}^{t'}K^{\alpha s}_{n'l'j'm_j'}\left(\zeta',\vec{x}\right)=\sum_{\lambda=0}^{2s-1}{}'\left[{}^{tt'}F^{\alpha s\lambda}_{nljm_j,n'l'j'm_j'}\left(\zeta,\zeta';\vec{x}\right)+{}^{tt'}\boldsymbol{F}^{\alpha s\lambda}_{nl_tjm_j,n'l_t'j'm_j'}\left(\zeta,\zeta';\vec{x}\right)\right] \tag{10a} $$

$$ \begin{aligned} {}^{tt'}F^{\alpha s\lambda}_{nljm_j,n'l'j'm_j'}\left(\zeta,\zeta';\vec{x}\right)=\eta_t\eta_{t'}\Big[&{}^{t}a^{ls}_{jm_j}(\lambda)\,{}^{t'}a^{l's}_{j'm_j'}(\lambda)\,k^{\alpha^*}_{nlm(\lambda)}(\zeta,\vec{x})k^{\alpha}_{n'l'm'(\lambda)}(\zeta',\vec{x})\\ &+{}^{t}a^{ls}_{jm_j}(\lambda+1)\,{}^{t'}a^{l's}_{j'm_j'}(\lambda+1)k^{\alpha^*}_{nlm(\lambda+1)}(\zeta,\vec{x})k^{\alpha}_{n'l'm'(\lambda+1)}(\zeta',\vec{x})\Big] \end{aligned} \tag{10b} $$

$$ \begin{aligned} {}^{tt'}\boldsymbol{F}^{\alpha s\lambda}_{nl_tjm_j,n'l_t'j'm_j'}\left(\zeta,\zeta';\vec{x}\right)=&\,{}^{t}a^{l_ts}_{jm_j}(2s-\lambda)\,{}^{t'}a^{l_t's}_{j'm_j'}(2s-\lambda)k^{\alpha^*}_{nl_tm(\lambda)}(\zeta,\vec{x})k^{\alpha}_{n'l_t'm'(\lambda)}(\zeta',\vec{x})\\ &+{}^{t}a^{l_ts}_{jm_j}(2s-(\lambda+1))\,{}^{t'}a^{l_t's}_{j'm_j'}(2s-(\lambda+1))k^{\alpha^*}_{nl_tm(\lambda+1)}(\zeta,\vec{x})k^{\alpha}_{n'l_t'm'(\lambda+1)}(\zeta',\vec{x}), \end{aligned} \tag{10c} $$

for STSO

$$ {}^{t}K^{s^+}_{nljm_j}\left(\zeta,\vec{x}\right){}^{t'}K^{s}_{n'l'j'm_j'}\left(\zeta',\vec{x}\right)=\sum_{\lambda=0}^{2s-1}{}'\left[{}^{tt'}F^{s\lambda}_{nljm_j,n'l'j'm_j'}\left(\zeta,\zeta';\vec{x}\right)+{}^{tt'}\boldsymbol{F}^{s\lambda}_{nl_tjm_j,n'l_t'j'm_j'}\left(\zeta,\zeta';\vec{x}\right)\right] \tag{11a} $$

$$ \begin{aligned} {}^{tt'}F^{s\lambda}_{nljm_j,n'l'j'm_j'}\left(\zeta,\zeta';\vec{x}\right)=\eta_t\eta_{t'}\Big[&{}^{t}a^{ls}_{jm_j}(\lambda)\,{}^{t'}a^{l's}_{j'm_j'}(\lambda)\,k^{*}_{nlm(\lambda)}(\zeta,\vec{x})k_{n'l'm'(\lambda)}(\zeta',\vec{x})\\ &+{}^{t}a^{ls}_{jm_j}(\lambda+1)\,{}^{t'}a^{l's}_{j'm_j'}(\lambda+1)k^{*}_{nlm(\lambda+1)}(\zeta,\vec{x})k_{n'l'm'(\lambda+1)}(\zeta',\vec{x})\Big] \end{aligned} \tag{11b} $$

$$
{}^{tt'}\boldsymbol{F}^{s\lambda}_{nl_t jm_j, n'l'_t j'm'_j}(\zeta,\zeta';\vec{x}) = {}^{t}a^{l_t s}_{jm_j}(2s-\lambda)\, {}^{t'}a^{l'_t s}_{j'm'_j}(2s-\lambda) k^{*}_{nl_t m(\lambda)}(\zeta,\vec{x}) k_{n'l'_t m'(\lambda)}(\zeta',\vec{x})
+ {}^{t}a^{l_t s}_{jm_j}(2s-(\lambda+1))\, {}^{t'}a^{l'_t s}_{j'm'_j}(2s-(\lambda+1)) k^{*}_{nl_t m(\lambda+1)}(\zeta,\vec{x}) k_{n'l'_t m'(\lambda+1)}(\zeta',\vec{x}). \tag{11c}
$$

ONE-RANGE ADDITION THEOREMS:

for ETSO

$$
{}^{t}K^{\alpha s\lambda}_{nljm_j}(\zeta,\vec{x}-\vec{y}) = \begin{bmatrix} \eta_t\, {}^{t}a^{ls}_{jm_j}(\lambda) k^{\alpha}_{nlm(\lambda)}(\zeta,\vec{x}-\vec{y}) \\ -\eta_t\, {}^{t}a^{ls}_{jm_j}(\lambda+1) k^{\alpha}_{nlm(\lambda+1)}(\zeta,\vec{x}-\vec{y}) \end{bmatrix} \tag{12a}
$$

$$
{}^{t}\boldsymbol{K}^{\alpha s\lambda}_{nl_t jm_j}(\zeta,\vec{x}-\vec{y}) = \begin{bmatrix} -i\, {}^{t}a^{l_t s}_{jm_j}(2s-\lambda) k^{\alpha}_{nl_t m(\lambda)}(\zeta,\vec{x}-\vec{y}) \\ -i\, {}^{t}a^{l_t s}_{jm_j}(2s-(\lambda+1)) k^{\alpha}_{nl_t m(\lambda+1)}(\zeta,\vec{x}-\vec{y}) \end{bmatrix}, \tag{12b}
$$

for STSO

$$
{}^{t}K^{s\lambda}_{nljm_j}(\zeta,\vec{x}-\vec{y}) = \begin{bmatrix} \eta_t\, {}^{t}a^{ls}_{jm_j}(\lambda) k_{nlm(\lambda)}(\zeta,\vec{x}-\vec{y}) \\ -\eta_t\, {}^{t}a^{ls}_{jm_j}(\lambda+1) k_{nlm(\lambda+1)}(\zeta,\vec{x}-\vec{y}) \end{bmatrix} \tag{13a}
$$

$$
{}^{t}\boldsymbol{K}^{s\lambda}_{nl_t jm_j}(\zeta,\vec{x}-\vec{y}) = \begin{bmatrix} -i\, {}^{t}a^{l_t s}_{jm_j}(2s-\lambda) k_{nl_t m(\lambda)}(\zeta,\vec{x}-\vec{y}) \\ -i\, {}^{t}a^{l_t s}_{jm_j}(2s-(\lambda+1)) k_{nl_t m(\lambda+1)}(\zeta,\vec{x}-\vec{y}) \end{bmatrix}, \tag{13b}
$$

where $\vec{x} \equiv \vec{r}, \vec{k}, \vec{\omega}_k$ and $\vec{y} \equiv \vec{R}, \vec{p}, \vec{\omega}_p$.

The formulas for the expansion and one-range addition theorems for quantities $\left(k^{\alpha^*}_{nlm}(\zeta,\vec{x}) k^{\alpha}_{n'l'm'}(\zeta',\vec{x}), k^{\alpha}_{nlm}(\zeta,\vec{x}-\vec{y})\right)$ and $\left(k^{*}_{nlm}(\zeta,\vec{x}) k_{n'l'm'}(\zeta',\vec{x}), k_{nlm}(\zeta,\vec{x}-\vec{y})\right)$ occurring on the right hand sides of these equations have been established in previous works [16, 17] and [18, 19], respectively.

As can be seen from the formulas of this work, all of the expansion an one-range addition theorems of 2(2s+1)-component ETSO and STSO defined in position, momentum and four-dimensional spaces are expressed through the corresponding nonrelativistic expansion and one-range addition theorems. Thus, the relations of nonrelativistic expansion and one-range addition theorems derived in previous papers [16-19] can be also used in the case of 2(2s+1)-component spinor orbitals in position, momentum and four-dimensional spaces.